\documentclass{acm_proc_article-sp}
\usepackage{epsfig}
\usepackage[square,comma,numbers,sort&compress]{natbib}
\usepackage{multirow}

\newcommand{\UPLB}{University of the Philippines Los Ba\~{n}os}

\begin{document}
\title{Microsimulations of Arching, Clogging, and\\Bursty Exit Phenomena in Crowd Dynamics}
\numberofauthors{1}
\author{
\alignauthor Francisco Enrique Vicente G. Castro  and Jaderick P. Pabico\\
   \affaddr{Institute of Computer Science}\\
   \affaddr{\UPLB}\\
   \affaddr{College 4031, Laguna}
   \email{\{fevgcastro, jppabico\}@uplb.edu.ph}
}
\date{}
\maketitle

\begin{abstract}
We present in this paper the behavior of an artificial agent who is a member of a crowd. The behavior is based on the social comparison theory, as well as the trajectory mapping towards an agent's goal considering the agent's field of vision. The crowd of artificial agents were able to exhibit arching, clogging, and bursty exit rates. We were also able to observe a new phenomenon we called double arching, which happens towards the end of the simulation, and whose onset is exhibited by a ``calm" density graph within the exit passage. The density graph is usually bursty at this area. Because of these exhibited phenomena, we can use these agents with high confidence to perform microsimulation studies for modeling the behavior of humans and objects in very realistic ways.
\end{abstract}

\section{Introduction}
Arching is a rainbow-like structure that naturally forms at the edge of a pedestrian crowd that jam and clog at exits. Clogging results as an effect of competition for space resource among members of a crowd who are unable to pass each other. Bursty exit rates are a result of jostling for position which prevents each pedestrian smooth passage along the exit width~\citep{henein05}. These phenomena are interesting to study and simulate because they are the most commonly observed behavior in crowd dynamics. In designing behavior for an agent in a multi-agent system (MAS), it is very important that the agents be able to exhibit these phenomena while in a crowd. When a crowd of artificial agents exhibits these phenomena, we then say that our microsimulation approach is more akin to modeling humans and real-world objects in very realistic ways. Thus, our microsimulation can be used with higher confidence to perform {\em what-if} scenarios to aid decision makers and researchers.

The growing number of students being admitted to various higher education institutions (HEI) in the country in the last 10 years, coupled with the slow-paced development of infrastructure to support the student population growth, particularly among the state colleges and universities (SUC), has resulted into crowding along building corridors that connect classrooms, lecture halls, and laboratories, despite the wishes of these institutions to provide better services and safer environments. Nowadays, students do not only contend for slots in subjects, which happen only once per semester during registration, but more so with walk spaces along the building corridors, which happen every class day. Students contend with corridor spaces whose nominal widths are effectively reduced because of the presence of other students, tables, chairs, and other pedestrian flow-retarding objects. Because the nominal widths of the corridors have become smaller than their potential widths, the effective travel times of students' egress along these corridors are lengthened, specifically during periodic flash crowd situations, such as during in-between class hours. If the walk-time of students along the building corridors is lengthened, specifically during normal egress conditions, how much more will be added to the evacuation time during chaotic panic situations (i.e., the faster-is-slower phenomenon) such as when there is a fire, or during an earthquake, or during riots or fraternity rumbles, or during critical incidents like chemical or biological spills? One need not wait for a disastrous event to answer this question because similar events can be simulated using a computer without compromising human lives.

\begin{figure*}[htb]
\centering\epsfig{file=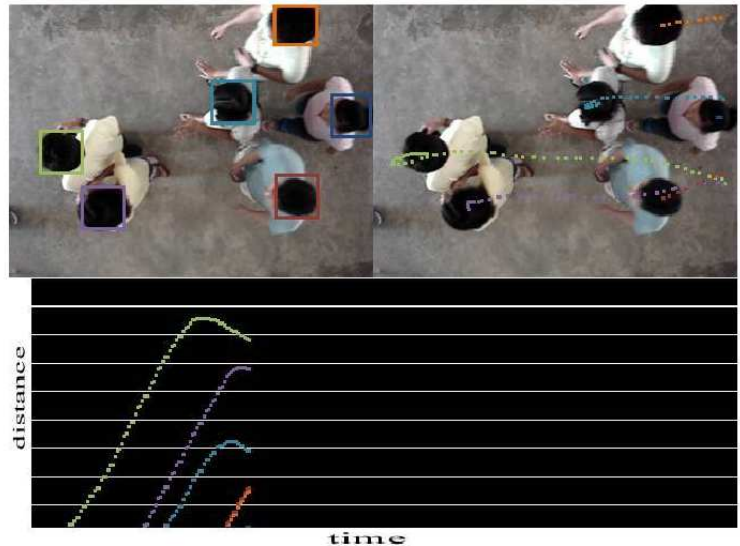, width=100mm}
\caption{An example output frame of the MVS. The upper Left part contains the input image of student pedestrians with trackers (bounding boxes). The upper Right part contains the trajectory of the pedestrians. The lower part contains the distance vs. time graph of the trajectory. This figure is in color in the electronic copy of this paper printed with permission from~\citet{ngoho09} and the Philippine Society of Information Technology Educators (PSITE).}\label{fig:1}
\end{figure*}

In this paper, we present our design of an artificial agent whose behavior resembles that of the humans when walking (or traveling) within a crowd. Our argument here is that if the crowd of our artificial agents can exhibit the phenomena that are usually observed in real-life, then we can use our agents in microsimulation studies of the dynamics of humans under normal egress and panic escape conditions. We present in this paper the behavior of our artificial agent, which is based on two recently developed theories: The social comparison theory~\citep{fridman10} and a trajectory mapping towards an agent's goal considering the agent's field of vision. With this behavior, our crowd of artificial agents was able to exhibit the arching, clogging, and bursty exit rates phenomena. The arch exhibited at the edge of the crowd resembles that of a half ellipse, with the major axis coincident along with the direction of crowd flow, while the minor axis is parallel to the exit width. We further present the effect of varying exit widths on the respective lengths of the major and minor axes in the arching phenomenon.

\section{Recent Advances in\\Crowd Dynamics}\label{sec:review}

In the past years, we developed a machine vision system (MVS) to automatically capture the dynamics of pedestrian students in a university building under four different traffic scenarios~\citep{ngoho09}. We considered the overhead view of each student as a digital object, where our MVS processes the image sequences to track the students. By considering the interactive effect of the camera lens perspective and the projected area of the corridor, the distance of each tracked student from its original position to its current position is approximated every video frame. The quantified motion of each tracked student are output by our MVS using 2-dimensional graphs of the kinematics of motion as a visualization (see for example Figure~\ref{fig:1}).

Because of the importance of crowd dynamics on several real-world applications, several researchers attempted to quantify the collective dynamics of the pedestrians through developed simulations of motion. \citet{helbing95}, borrowing some ideas in gas-kinematic models, introduced the social force model (SFM) to simulate pedestrian flows.  In their model, a self-driven particle (i.e., a pedestrian) that interacts through social rules and regulations tries to move in its desired speed and direction while at the same time attempts not to collide with obstacles, other particles, and surrounding barriers.  In order to reach its destination faster, pedestrians take detours even if the route is crowded~\citep{helbing01}. The choice, however, is dependent on the recent memory of what the traffic was like the last time they took the route, the figure of which was found by other researchers to be polygonal in nature~\citep{ganem98}. In agreement with the social force model, ~\citet{weidmann93} observed that, as long as it is not necessary to go faster, such as going down a ramp, a pedestrian prefers to walk with his or her desired speed, which corresponds to the most comfortable walking speed. However, ~\citet{weidmann93} further observed that pedestrians keep a certain distance from other pedestrians and borders. The distance between the pedestrians decreases as the density of the crowd increases. The pedestrians themselves cause delays and obstructions. \citet{arns93} observed that the motion of the crowd is similar to the motion of gases and fluid, while~\citet{helbing01} suggested that it is similar to granular flow as well. 

~\citet{helbing01}, in his extension of the SFM, showed that many aspects of traffic flow can be reflected by self-driven many-particle systems. In this system, he identified the various factors that govern the dynamics of the particles such as the specification of the desired velocities and directions of motion, the geometry of the boundary profiles, the heterogeneity among the particles, and the degree of fluctuations. One such observable pattern is the formation of lanes of uniform walking direction, formed because of the self-organization of the pedestrians~\citep{helbing06}. Aside from the self-organizing behavior of the crowd, obstacles were also observed to both positively and negatively contribute to the flow of the traffic. 

\begin{figure*}[hbt]
\centering\epsfig{file=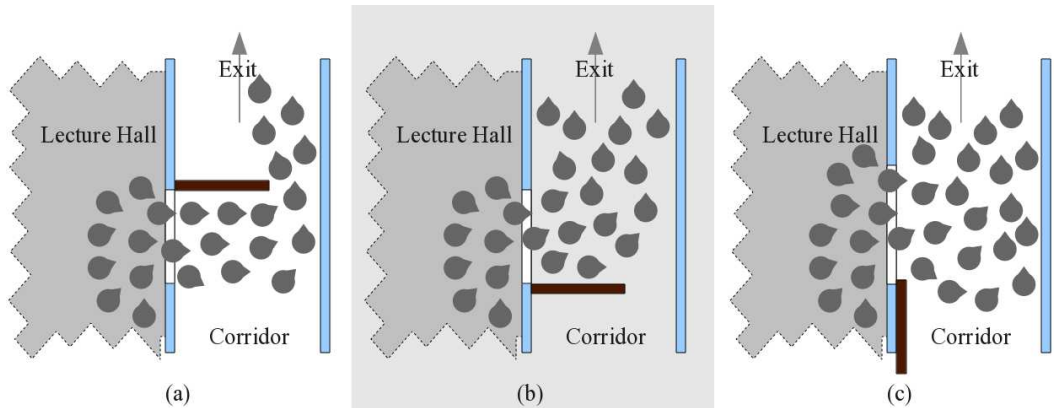, width=140mm}
\caption{Snapshots of an example {\em what-if} scenario simulations on the effect of exit door configurations to student egress in a very large lecture hall (students are shown as black circles with pointers to visualize heading):  (a) Current exit door configuration which opens outward but hinders student flow towards the corridor exit; (b) Effect of moving the door hinge such that the exit door opens behind the student flow; and (c) Effect of increasing the effective width of the exit door by 50\% and replacing the swinging door by a sliding door and notice the effect on the utilization of the nominal corridor width. This figure is in color in the electronic copy of this paper. The artwork is licensed under a Creative Commons Attribution-NonCommercial-NoDerivs 3.0 Unported License.}\label{fig:2}
\end{figure*}

During escape panic of large crowds, several behavioral phenomena were observed~\citep{helbing00}: build up of pressure, clogging effects at bottlenecks, jamming at room widening areas, faster-is-slower effect, inefficient use of alternative exits, initiation of panics by counter flows, and impatience. It was observed that the main contributing behaviors in these situations is a mixture of individual and grouping behavior. 

\section{Methodology}\label{sec:method}

\subsection{General Behavior}
In our simulation, each agent was provided with the following general behaviors described originally by~\citet{wooldridge95}, and later on extended and explained further by~\citet{frankling96,epstein99,torrens04} and~\citet{macal05}:

\begin{enumerate}
\item {\em Autonomy}: We program our agents as autonomous units (i.e. governed without the influence of centralized control). Each agent is capable of processing information and exchanging this information with other agents in order to make independent decisions. They are free to interact with other agents, at least over a limited range of situations, and this does not (necessarily) affect their autonomy. In this respect, we say that our agents are active rather than purely passive (see item number~\ref{item:Active} below). 
\item {\em Heterogeneity}: We believe that the notion of mean-individuals is redundant. Instead we believe that our agent's programming permit the development of autonomous individuals. When these agents are put into a group, such as in a crowd, the formed group is a heterogenous one composed of different yet similarly-programmed agents. Therefore, groups of agents can exist, but programmatically the groups are spawned from the bottom-up, and can be clearly seen as amalgamations of similarly-programmed autonomous individuals. 
\item {\em Active}\label{item:Active}: Our agents are active because they exert independent influence in a simulation. For example, an agent can directly affect the decision made by other agents, or its very presence in the simulated environment can greatly affect changes in that environment and thus indirectly affect other agents. Because of this, we can identify the following active features of our agents: 
  \begin{enumerate}
  \item Pro-active or goal-directed
  \item Reactive or perceptive
  \item Bounded rationality
  \item Interactive and communicative
  \end{enumerate}
\end{enumerate}

Because our agents exhibit these behaviors, our simulation approach is more akin to modeling humans and objects in very realistic ways than other modeling approaches, such as those that aggregate mathematical equations to explain the dynamics of pedestrian, say, during panic situations or emergency evacuations. Thus, our simulation can be used with higher confidence to perform what-if scenarios to aid, for example, university administrators and decision makers with regards to management policies, as well as infrastructure development plans, for safer learning environments for the students and constituents (see for example Figure~\ref{fig:2}).

\begin{figure}[htb]
\centering\epsfig{file=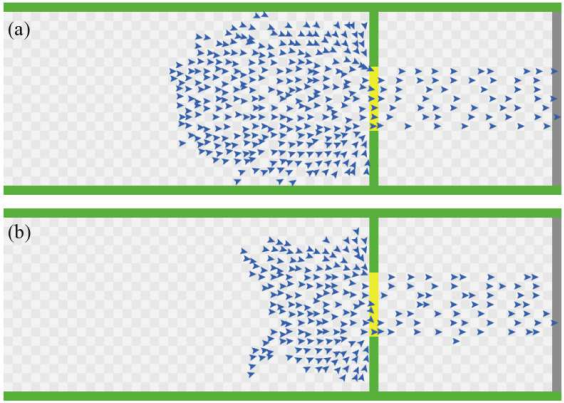, width=75mm}
\caption{Profile of the crowd before the exit door: (a) Onset of arching where the minor and major axes are visible; and (b) Occurrence of double arches toward the end of the simulation.}\label{fig:arching}
\end{figure}

\begin{figure}
\centering\epsfig{file=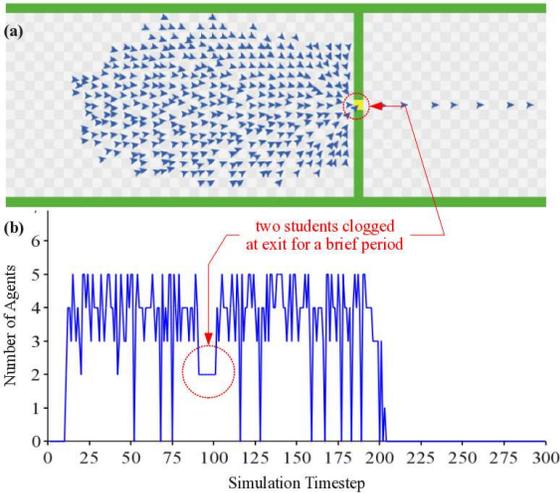, width=75mm}
\caption{(a) Profile of the crowd at a clogged exit door. (b) Density graph at the exit door shows that no agent was able to pass the exit during the clogging effect.}\label{fig:clogging}
\end{figure}

\subsection{Agent Behavior}
In a two-dimensional flat world, let~$G$ be the goal of all agents with set coordinates $\{(x_1, y_1), (x_2, y_2), \dots, (x_n, y_n)\}$. This set is usually the location of an exit door. Since all agents aim to reach any one of the exit coordinates, the agents will face towards the direction of the nearest exit coordinates from their current coordinates. The agent will scan its field of vision for the closest free space and move towards that space with its gaiting speed, which, in this paper, is currently set at one pace per simulation time step. In the future, we wish to vary this gaiting speed depending on the agent's height. A free space is a location in the flat world that is not a wall, another agent, or any other object. If other agents are blocking the agent's space within its field of vision, the agent stops at that coordinates. It is possible that the agent might move away from the target exit coordinate if the chosen free space was at the edge of its field of vision. When that happens, the agent will still move to the free space, but after moving, it will redirect its heading towards the possibly new nearest exit coordinates. When the distance of the agent to the nearest exit coordinates is $<1$, it considers itself as already exited and will move to the edge of the simulation world. Until this distance has not been achieved, the agent will just repeat the agent's decision making process described above.

\subsection{Effect of Exit Width on\\Arching Profile}

We conducted experiments using a crowd of 400 agents to see the effect of various widths of the exit door to the length of the arch's minor and major axes. Based on our preliminary simulations, 400 was the minimum number of agents that can exhibit arching, given each agent's gait speed and a corridor width of 19. The width is in terms of agent's width and the various exit widths that we used were: 1, 3, 5, 7, 9, 11, and 13. The various exit widths were computed by adding one person width on each side of the exit along the minor axis.

\section{Results and Discussion}\label{sec:results}

\subsection{Arching Effects}

Figure~\ref{fig:arching}a shows the profile of a crowd that exhibits arching near the exit door. We observed that the arch resembles a half ellipse with the ellipse's major axis running along the direction of the crowd flow and perpendicular to the exit width, while the minor axis is parallel to the exit width. This result confirms that of the other researchers'~\citep{frankling96,epstein99,helbing01}. One particular result that we feel contributed new knowledge on this field is our observation of the occurrence of what we called ``double arches'' that formed towards the end of the simulation (Figure~\ref{fig:arching}b). As far as we are concerned, none of the literature that we reviewed described this occurrence. The double arches appeared when the clogging at the exit loosened up, and more agents that are near and along the major axis were able to pass through. We also observed that the burstiness of the exit rate disappeared at the start of the formation of double arches. We described this disappearance of burstiness as ``calm'' in the visualization of the phenomenon using the density graph at the exit area.

\subsection{Clogging}

Clogging results when two or more agents compete for the same corridor space and not one of them are able to pass each other. Figure~\ref{fig:clogging}a shows a snap-shot of the simulation where clogging occurred. The accompanying density graph at the exit door (Figure~\ref{fig:clogging}b) confirms that no agent was able to pass the exit. As opposed to the event of a bursty exit rate where agents temporarily clog the exit door but due to emergent behavior are able to unclog themselves, clogging occurred because the agents were not able to resolve their conflict in grabbing the space resource just in front of them.

\subsection{Bursty Exit Rate}

\begin{figure}
\centering\epsfig{file=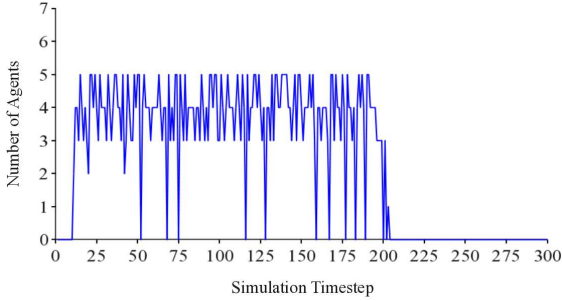, width=75mm}
\caption{Density graph at the exit door shows a bursty exit rate.}\label{fig:bursty}
\end{figure}

Figure~\ref{fig:bursty} shows the density graph of the crowd at the exit door. The saw-tooth-like form of the density graph confirms that all exit scenario exhibit a bursty rate. The burstiness can be attributed to agents competing for position which for a brief moment clogged the exit way. However, through the agents' emergent behavior, they were able to unclog themselves, and then pass the exit door in a burst carrying more agents. This results to abrupt change of density at the exit area, and is visualized as a saw-tooth-like line in the density graph.

\subsection{Effect of Exit Width on\\Arching Profile}

\begin{figure}[bh]
\centering\epsfig{file=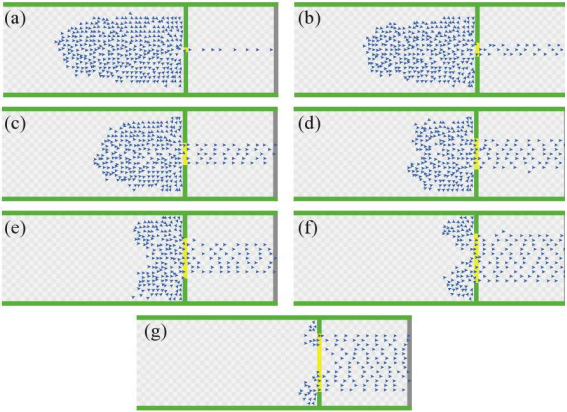, width=75mm}
\caption{Arch profile at various door widths: (a) 1; (b) 3; (c) 5; (d) 7; (e) 9; (f) 11; and (g) 13. }\label{fig:width}
\end{figure}

Figure~\ref{fig:width} shows the effect of varying the width of the exit door on arching profile. We can see here that at narrower exit widths, the major axis is longer, while the minor axis is shorter. When the exit door is wider, the major axis is shorter than normal, while the minor axis is longer than normal. 

\section{Summary and Conclusion}\label{sec:conclude}

In this paper, we present the behavior for an artificial agent used in microsimulation of pedestrian crowd. The behavior is based on the two recently developed theories, the social comparison theory~\cite{fridman10} and a trajectory mapping towards an agent's goal considering the agent's field of vision. With these behaviors, we were able to show that the crowd with these agents was able to exhibit arching, clogging, and bursty exit rate. We also observed a new phenomenon that occurred towards the end of the simulation that we called Double Arching. As far as our literature review is concerned, we are the first to describe the emergent behavior of double arching in the crowd dynamics, which happen during the temporary ``calm" that occurred in the bursty exit rate graph. We also described the behavior of the density graph during clogging, and described clogging to have occurred when two or more agents compete for space near the exit. We have also shown that the property of bursty exit rate is exhibited by the density graph with a saw-tooth-like profile. Lastly, we conducted an experiment on the effect of various exit door widths on the arching profile of the crowd before the exit door. At a narrow exit width, the major axis of the arch tends to be longer, while the minor axis tends to be shorter. However, at a wide exit width the major axis of the arch tends to be shorter, while the minor axis tends to be longer. Because our crowd of agents were able to exhibit the arching, clogging, and bursty exit rate phenomena in crowd dynamics, we can conclude that we can use these agents in microsimulation studies for modeling the behavior of humans and objects in very realistic ways. Thus, we can use these microsimulations with higher confidence to perform what-if scenarios to aid in decision making.

\section{Extension}

The following efforts are already underway as extensions to this research endeavor:
\begin{enumerate}
\item As we explained earlier, the arching phenomenon is an emergent pattern formed by a $c$-sized crowd of intelligent, goal-oriented, autonomous, heterogeneous individuals moving towards a $w$-wide exit along a $W$-wide corridor, where $W>w$. We are currently collecting empirical data from microsimulations to identify the combination effects of~$c$ and~$w$ to the time $T$ of the onset of and the size~$S$ of the formation of the arch. We aim to measure the~$S$ with respect to the lengths of the major and minor axes of the ellipse, respectively.
\item The total time~$T$ of egress of large crowds, such as students exiting a large lecture hall filled to its maximum capacity, is hindered by the number of exit doors, as well as the doors' dimensions, positions, and orientation. In this current endeavour, we aim to find out the combinatorial effects of the number of exit doors, their dimensions, and their orientations using microsimulations of our agents. For our simulations to be realistic, we will scale down a real physical structure of a large lecture hall, including the dimensions of its exit doors, into a two-dimensional simulation. We plan to fill the simulated lecture hall with agents corresponding to the full capacity of the real lecture hall. We will then conduct experiments on various crowd sizes~$C$ and different exit door scenarios~$D$. We will then measure~$T$ under all $C \times D$ combinations.
\item We would like to compare the capabilities of the SFM and our approach in simulating the following real-world crowd phenomena: ``faster-is-slower'' in escape panic, ``arching'' and ``bursty exit'' as side effects to ``clogging'' on exit ways, ``flocking,'' ``bidirectional lane formation,'' and ``roundabout formation.'' We believe that our approach is also able to exhibit two more individual behaviors that the SFM can not do: (1) {\em Imitation} -- where individuals tend to move into groups whose members they thought would have the same opinion as theirs; and (2) {\em Contagion} -- where people tend to ``adopt'' the behavior of others in the same group.
\end{enumerate}

\section*{Acknowledgment}
This research effort is under the research program {\bf Multi-agent Simulation of Student Egress from the ICS Mega Lecture Hall Under Normal, Controlled Emergency, and Panic Situations} funded by the {\em Institute of Computer Science, \UPLB}.
\bibliography{microsimulations}

\begin{thebibliography}{15}
\providecommand{\natexlab}[1]{#1}
\providecommand{\url}[1]{\texttt{#1}}
\expandafter\ifx\csname urlstyle\endcsname\relax
  \providecommand{\doi}[1]{doi: #1}\else
  \providecommand{\doi}{doi: \begingroup \urlstyle{rm}\Url}\fi

\bibitem[Arns(1993)]{arns93}
T.~Arns.
\newblock Video films of pedestrian crowds, 1993.
\newblock Wannenstr. 22, 70199 Stuttgart, Germany.

\bibitem[Epstein(1999)]{epstein99}
J.M. Epstein.
\newblock Agent-based computational models and generative social science.
\newblock \emph{Complexity}, 4\penalty0 (5):\penalty0 41--60, 1999.

\bibitem[Franklin and Graesser(1996)]{frankling96}
S.~Franklin and A.~Graesser.
\newblock Is it an agent, or just a program? {A} taxonomy for autonomous
  agents.
\newblock In \emph{Proceedings of the Third International Workshop on Agent
  Theories, Architectures, and Languages}. Springer-Verlag, 1996.

\bibitem[Fridman and Kaminka(2010)]{fridman10}
N.~Fridman and G.A. Kaminka.
\newblock Modeling pedestrian crowd behavior based on a cognitive model of
  {S}ocial {C}omparison {T}heory.
\newblock \emph{Computational and Mathematical Organization Theory},
  16:\penalty0 348--372, 2010.

\bibitem[Ganem(1998)]{ganem98}
J.~Ganem.
\newblock A behavioral demonstration of {F}ermat's principle.
\newblock \emph{The Physics Teacher}, 36, 1998.

\bibitem[Helbing(2006)]{helbing06}
D.~Helbing.
\newblock \emph{Safety Management at Large Events: {T}he Problem of Crowd
  Panic}.
\newblock Institute for Transport and Economics, Dresden University of
  Technology, 2006.

\bibitem[Helbing and Molnar(1995)]{helbing95}
D.~Helbing and P.~Molnar.
\newblock Social force model for pedestrian dynamics.
\newblock \emph{Physical Review E}, 51\penalty0 (5):\penalty0 4282+, 1995.

\bibitem[Helbing et~al.(2000)Helbing, Farkas, and Vicsek]{helbing00}
D.~Helbing, I.~Farkas, and T.~Vicsek.
\newblock Simulating dynamical features of escape panic.
\newblock \emph{Nature}, 407\penalty0 (6803):\penalty0 487--490, 2000.

\bibitem[Helbing et~al.(2001)Helbing, Molnar, Farkas, and Bolay]{helbing01}
D.~Helbing, P.~Molnar, I.~Farkas, and K.~Bolay.
\newblock Self-organizing pedestrian movement.
\newblock \emph{Environment and Planning B: Planning and Design}, 28:\penalty0
  361--383, 2001.

\bibitem[Henein and White(2005)]{henein05}
C.M. Henein and T.~White.
\newblock \emph{Agent-based Modeling of Forces in Crowds}, volume 3415, pages
  173--184.
\newblock 2005.

\bibitem[Macal and North(2005)]{macal05}
C.M. Macal and M.J. North.
\newblock Tutorial on agent-based modelling and simulation.
\newblock In M.E. Euhl, N.M. Steiger, F.B. Armstrong, and J.A. Joines, editors,
  \emph{Proceedings of the 2005 Winter Simulation Conference}, 2005.

\bibitem[Ngoho and Pabico(2009)]{ngoho09}
L.V.A. Ngoho and J.P. Pabico.
\newblock Capturing the dynamics of pedestrian traffic using a machine vision
  system.
\newblock \emph{Philippine Information Technology Journal}, 2\penalty0
  (2):\penalty0 1--11, 2009.

\bibitem[Torrens(2004)]{torrens04}
P.M. Torrens.
\newblock \emph{Simulating Sprawl: {A} Dynamic Entity-Based Approach to
  Modelling {N}orth {A}merican Suburban Sprawl Using Cellular Automata and
  Multi-Agent Systems}.
\newblock PhD thesis, University College London, London, 2004.

\bibitem[Weidmann(1993)]{weidmann93}
U.~Weidmann.
\newblock Transportation technique for pedestrians, 1993.
\newblock Zurich, Switzerland.

\bibitem[Wooldridge and Jennings(1995)]{wooldridge95}
M.~Wooldridge and N.R. Jennings.
\newblock Intelligent agents: {T}heory and practice.
\newblock \emph{Knowledge Engineering Review}, 10\penalty0 (2):\penalty0
  115--152, 1995.

\end{thebibliography}
\bibliographystyle{plainnat}

\end{document}